\documentstyle[12pt]{article}
\newcommand{\ms}{m_{{\scriptscriptstyle S}}}

\newcommand{\be}{\begin{equation}}
\newcommand{\ee}{\end{equation}}
\newcommand{\bea}{\begin{eqnarray}}
\newcommand{\eea}{\end{eqnarray}}
\newcommand{\beastar}{\begin{eqnarray*}}
\newcommand{\eeastar}{\end{eqnarray*}}

\renewcommand{\;}{\ ;\ \ \ \ }
\newcommand{\refe}[1]{(\ref{#1})}
\newcommand{\mm}{\hspace{0.5mm}}

\newcommand{\eq}{Equation }

\newcommand{\half}{{\textstyle \frac{1}{2}}}
\newcommand{\tilh}{\tilde h}
\renewcommand{\t}{t}
\newcommand{\Z}{{\cal Z}}
\newcommand{\N}{{\cal N}}
\newcommand{\abs}[1]{\left| #1 \right|}
\newcommand{\e}[1]{\ {\rm e}^{ #1 }\mm}
\newcommand{\mat}[2]{\left(\begin{array}{#1} #2 \end{array}\right)}
\newcommand{\stack}[2]{\stackrel{{\scriptscriptstyle #1}}{#2}}
\renewcommand{\cases}[1]{\left\{\begin{array}{cl} #1\end{array}\right.}

\begin{document}
\baselineskip 17pt plus 2pt minus 1pt
\pagestyle{empty}
\begin{flushright}
{ DFUPG-109-95\\
hep-th/9511110}
\end{flushright}


\begin{center}
\vspace {2. cm}

{\bf Charge Screening in the Finite Temperature Schwinger Model}\footnote{ This
work is supported in part by the Istituto Nazionale di Fisica Nucleare of
Italy,
the Natural Sciences and Engineering Research Council of Canada and a NATO
Scientific Exchange Grant}\\
\vspace {1. cm}
{\large G. Grignani$^a$, G. Semenoff$^b$, P. Sodano$^a$ and O.
Tirkkonen$^b$}\\
\vspace {0.4cm}
$(a)$Dipartimento di Fisica\\ and Sezione I.N.F.N.\\Universit\'a di
Perugia\\ Via a Pascoli I-06100, Perugia, Italia\\~~~\\
$(b)$Department of Physics,\\ University of British
Columbia\\Vancouver, British Columbia,\\ Canada V6T 1Z1
\\
\vspace {0.4cm}
\end{center}
\vspace {1 cm}
\centerline{\bf Abstract}
\vspace {0.4 cm}

\noindent
We compute the effective action and correlators of the Polyakov loop
operator in the Schwinger model at finite temperature and discuss the
realization of the discrete symmetries that occur there. We show that,
due to nonlocal effects of massless fermions in two spacetime
dimensions, the discrete symmetry which governs the screening of
charges is spontaneously broken even in an effective one-dimensional
model, when the volume is infinite.  In this limit, the thermal state
of the Schwinger model screens an arbitrary external charge;
consequently the model is in the deconfined phase, with the charge of
the deconfined fermions completely screened.  In a finite volume we
show that the Schwinger model is always confining.


\newpage
\pagestyle{plain}

\section{Introduction}

The Schwinger model ~\cite{sch} is an exactly solvable model which
describes the electrodynamics of a massless fermion in 1+1 spacetime
dimensions.  It is the classic example of a confining gauge theory and
displays some of the features of quantum chromodynamics or other
higher dimensional gauge theories where strong infrared effects are
important ~\cite{cks}.  In one space dimension the tree level
electron-positron potential, $V(x)\propto {\rm e}^2\vert x\vert$, is already
confining without quantum fluctuations.  Detailed analysis in both the
path integral ~\cite{lsb,bz} and operator methods
{}~\cite{ls,cf1,cf2} shows that the spectrum is completely gapped,
exhibits chiral symmetry breaking and has no charged excitations.

Over the past few decades, there has been considerable interest in the
properties of the Schwinger model at finite temperature, both in the
massless {}~\cite{l,a,sv,ra,d,tz,bp,sw,k,smilga,fhnvz,hnz,ssz} and in
the massive ~\cite{hhi} case.  The most interesting question as to
whether the chiral symmetry breaking seen in the vacuum is restored at
high temperatures was answered in the negative long ago {}~\cite{dj}.
The breaking of the chiral symmetry is a consequence of the axial
anomaly, rather than spontaneous symmetry breaking, and the axial
anomaly, being a result of short distance physics, is insensitive to
temperature.  Thus, many of the features of the Schwinger model are
not changed by temperature.

In this paper, we advocate the use of temperature to explore the
spectrum of the Schwinger model.  The thermal state is a density
matrix with non-vanishing contributions from all states in the
spectrum with finite energy and thus contains information about all of
the states.

As an infrared regularization, we shall consider the space as a circle
of circumference $L$ on which all of the basic fields of the Schwinger
model have periodic boundary conditions.  The Hamiltonian quantization
in this regularization and some of the questions concerning topology
and theta-vacua which arise in this case were discussed by Manton
{}~\cite{m}. These issues, as well as questions of Wilson line phase
dynamics and correlators, were further developed in Ref.
\cite{hethos}.

Part of our motivation for this work is to test a recent idea
{}~\cite{gss} that the Polyakov loop operator, introduced by Polyakov
{}~\cite{poly} and Susskind ~\cite{suss} as an order parameter for
confinement in non-abelian Yang-Mills theory in higher dimensions is
also a useful operator for Abelian gauge theory.  In non-Abelian gauge
theory, the Polyakov loop has the limitation that it is an order
parameter for confinement only in models where all of the fields are
invariant under global gauge tranforms in the center of the gauge
group, i.e. are in the adjoint or some other zero `N-ality'
representation.  In electrodynamics, on the other hand, it can be used
in any model which is essentially compact in the sense that all of the
charges of the dynamical fields are integer multiples of some basic
charge ~\cite{gss}.  Then, as was argued in ~\cite{gss}, the Polyakov
loop with an incommensurate charge can be used to probe the ability of
the electrodynamic system to screen external charges.

\subsection{$Z_N$ symmetry of finite temperature Yang-Mills theory}

We shall first review the role of the Polyakov loop operator as an
order parameter for confinement in Yang-Mills theory at finite
temperature.  This is conventionally seen in the Euclidean path
integral formulation of the finite temperature gauge theory.  In that
formulation, the Polyakov loop operator measures the holonomy of the
gauge connection in the periodic Euclidean time,
\begin{equation}
P(\vec x)\equiv {\rm tr}{\cal P}\exp\left(i\int_0^{1/T}d\tau A_0(\tau,
\vec x)\right)
\label{ploop}
\end{equation}
whose correlators in a finite temperature Yang-Mills theory are defined
by the Euclidean path integral
\bea
\langle P(\vec x_1)\ldots P(\vec x_m)P^{\dagger}(\vec y_1)\ldots
P^{\dagger} (\vec y_n)\rangle \hskip 60mm \cr
\cr
\qquad\qquad  = \frac{ \int dA_\mu e^{-\int{\rm
tr}F^2/4} P(\vec x_1)\ldots P(\vec x_m)P^{\dagger}(\vec y_1)\ldots
P^{\dagger} (\vec y_n)}{ \int dA_\mu e^{-\int_0^{1/T}{\rm tr}F^2/4} }
\eea
where the gauge field has periodic boundary conditions,
\begin{equation}
A_\mu(1/T,\vec x)=A_\mu(0,\vec x)
\end{equation}
Since the Yang-Mills field transforms in the adjoint representation
of the gauge group,
\begin{equation}
A_\mu'(\tau,\vec x)=g^{-1}(\tau,\vec x)A_\mu(\tau,\vec x)g(\tau,\vec x)
+ig^{-1}(\tau,\vec x)\nabla_\mu(\tau,\vec x)g(\tau,\vec x)
\end{equation}
they remain periodic under gauge transformations which are periodic
up to an element, $Z$,  of the center of the group,
\begin{equation}
g(1/T,\vec x)=Z g(0,\vec x)
\label{pgt}
\end{equation}
The center of $SU(N)$ is $Z_N$, the additive group of the integers modulo
$N$, whereas the center of $U(N)$ is the Abelian group $U(1)$.
The coset of the group of all gauge transformations modulo those which
are strictly periodic is a global transformation by elements
in the center of the gauge group.

As well as pure Yang-Mills theory, any gauge theory which has matter
fields which transform in the adjoint, or any other zero `N-ality'
representation of the gauge group will have this symmetry of the path
integral. Furthermore, this symmetry exists for any gauge group which
has a non-trivial center.  In the following, we shall take the gauge
group to be SU(N), with the center $Z_N$. The non-trivial topological
structure arises from the non-vanishing first homotopy
$\Pi_1(SU(N)/Z_N) = Z_N$. An important question is whether or not this
$Z_N$ symmetry is spontaneously broken at finite temperature.

Under the gauge transformation (\ref{pgt}),
the Polyakov loop operator transforms as
\begin{equation}
P'(\vec x)=Z P(\vec x)
\end{equation}
Therefore, this operator can be used
as an order parameter for breaking of the $Z_N$ symmetry.

The connection between the breaking of $Z_N$ symmetry and
confinement is through the fact that the
correlators of Polyakov loop operators
\begin{equation}
e^{-F(\vec x_1,\ldots,\vec x_m,\vec y_1,\ldots,\vec y_n)/T}
=\langle P(\vec x_1)\ldots P(\vec x_m)P^{\dagger}
(\vec y_1)\ldots P^{\dagger}(\vec y_n)\rangle
\label{cor}
\end{equation}
can be interpreted as giving the free energy $F(\vec x_1,\ldots,
\vec x_m,\vec y_1,\ldots,\vec y_n)$ of the finite
temperature gauge theory with an array of classical, external,
fundamental representation quark sources at positions $\vec x_1,
\ldots,\vec x_m$ and anti-quark sources at positions $\vec y_1,
\ldots, \vec y_n$.  The normalization of the correlator subtracts the
free energy of the gauge theory at the same temperature
in the absence of sources.

If the $Z_N$ symmetry is not spontaneously broken, the correlator in
(\ref{cor}) vanishes unless $m=n$ modulo $N$, i.e.  unless the quarks
and anti-quarks occur in the right numbers to make up mesons, which
are quark-antiquark pairs, or baryons or anti-baryons, which are
groups of $N$ quarks or $N$ anti-quarks, respectively.  The vanishing
of the correlator is interpreted as the quark charge distribution
having infinite free energy when it has quantum numbers which cannot
be combined into color singlets, i.e. as quark confinement.  On the
other hand, if the $Z_N$ symmetry is spontaneously broken, the
correlators can be non-zero and even quark charge distributions which
cannot form color singlets can have finite free energy.

Furthermore, the free energy $F(\vec x_1,\ldots,\vec x_m,\vec y_1,
\ldots,\vec y_n)$ can be viewed as the effective potential energy
of the array of quarks and antiquarks.  For example, the effective
interaction between a quark and anti-quark is given by
$F(\vec x,\vec y)$.  If this correlator increases with distance, as
it would in a confined phase where there is a non-zero string tension,
then the correlator of Polyakov loop operators would have the clustering
property
\be
\lim_{\vert x-y\vert\rightarrow\infty}
e^{-F(\vec x,\vec y)/T}=
\lim_{\vert x-y\vert\rightarrow\infty}
\langle P(\vec x)P^{\dagger}(\vec y)\rangle =0
\ee
This implies that the $Z_N$ symmetry is unbroken and is consistent with
the vanishing of the expectation value of the loop operator,
\be
\langle P(\vec x)\rangle =0
\ee
On the other hand, in the deconfined phase, one would expect the
quark-antiquark potential to fall off to zero with some screening
length (the non-Abelian analog of Debye screening).  In that case,
\be
\lim_{\vert x-y\vert\rightarrow\infty}
e^{-F(\vec x,\vec y)/T}=\lim_{\vert x-y\vert\rightarrow\infty}
\langle P(\vec x)P^{\dagger}(\vec y)\rangle\neq 0
\ee
This implies that the $Z_N$ symmetry is spontaneously
broken and is consistent with the loop operator having a non-vanishing
expectation value
\be
\langle P(\vec x)\rangle\neq 0
\ee
To properly compute this one-point function, one should as usual introduce
a symmetry breaking external field through a term in the action
such as
\be
S_\lambda=\lambda~\int d\vec x ~\left( P(\vec x)+P^{\dagger}(\vec x)\right)
\ee
and compute the one-point function, which would be non-zero when $\lambda$ is
not zero.  Then, the occurence of symmetry breaking would be seen when the
limit
\be
\lim_{\lambda\rightarrow 0}\langle P(\vec x)\rangle_\lambda
\ee
is non-zero.

This formalism is well developed for finite temperature Yang-Mills theory
and some other pure gauge theories such as compact $U(1)$  and some $Z_N$
gauge theory.  All non-trivial pure gauge theories in spacetime dimensions
greater than two exhibit a high-temperature
de-confined phase and almost all have a phase transition to a confined phase
at some critical temperature.
Details are summarized in the comprehensive review by
Svetitsky, ~\cite{bs}.

However, for gauge theories with dynamical quarks, the
Polyakov loop operator is not a useful order parameter to characterize
a confining phase.  The reason is that, since the quark fields transform
in the fundamental representation of the gauge group, their action and
boundary conditions in the path integral are invariant under only strictly
periodic gauge transformations.  Thus, fundamental representation quarks
are said to break the $Z_N$ symmetry explicitly.  The free energy of
a distribution of external quarks is always finite.  This is interpreted as
the possibility of pair creation of dynamical quark-antiquark pairs so that
the dynamical quarks can screen the color of any external distribution
of quark or anti-quark sources.

\subsection{$Z$ symmetry of quantum electrodynamics}

In an Abelian gauge field theory, the Polyakov loop operator is defined by
the analog of (\ref{ploop})
\begin{equation}
P(\vec x)=\exp\left( i\int_0^{1/T}d\tau A_0(\tau,x)\right)
\end{equation}
and its correlators are computed using the
finite temperature path integral of quantum electrodynamics
\begin{equation}
\langle \prod_i P_{e_i}(\vec x_i)\rangle =
\frac{ \int dA_\mu d\psi d\bar\psi e^{-\int_0^{1/T} F_{\mu\nu}^2/4+\bar\psi
(\gamma\cdot D+m)\psi} \prod_i e^{ie_i\int_0^{1/T}d\tau A_0(\tau,\vec x_i)} }
{  \int dA_\mu d\psi d\bar\psi e^{-\int_0^{1/T} F_{\mu\nu}^2/4+\bar\psi
(\gamma\cdot D+m)\psi } }
\end{equation}
with the (anti-)periodic boundary conditions,
\begin{equation}
A_\mu(1/T,\vec x)=A_\mu(0,\vec x)
\end{equation}
\begin{equation}
\psi(1/T,\vec x)=-\psi(1/T,\vec x) ~~,~\bar\psi(1/T,\vec x)=-\bar\psi(0,\vec x)
\end{equation}
These boundary conditions, as well as the measure
and action in the functional integral are invariant
under the gauge transformation
\begin{equation}
A_\mu '(\tau,\vec x)=A_\mu(\tau,\vec x)+\nabla_\mu\chi(\tau,\vec x)
\end{equation}
\begin{equation}
\psi'(\tau,\vec x)=e^{ie\chi(\tau,\vec x)}\psi(\tau,x)
{}~~,~
\bar\psi'(\tau,\vec x)=e^{-ie\chi(\tau,\vec x)}\bar\psi(\tau,x)
\end{equation}
when the gauge function is periodic up to an additive constant
\begin{equation}
\chi(1/T,\vec x)=\chi(0,\vec x)+2\pi n/e
\end{equation}
where $n$ is an integer.  The coset of the group of all allowed gauge
transformations modulo the group of all strictly periodic gauge
transformations is $\Pi_1(U(1)) = Z$, the additive group of the
integers.  Note that this is a symmetry of the functional integral
representation of the partition function even in the presence of
dynamical electrons.  It is an interesting and well-defined question
to ask whether this symmetry is realized in a spontaneously broken or
unbroken phase in quantum electrodynamics.

The Abelian Polyakov loop operator
transforms under a gauge transformation as
\begin{equation}
P_{\tilde e}'(\vec x)= e^{2\pi n\tilde e/e}P_{\tilde e}(\vec x)
\end{equation}
and is not invariant under the $Z$ symmetry
unless $\tilde e$ is an integer multiple of the electron charge $e$.  This
transformation law was noted by Hansson, Nielsen and Zahed \cite{hnz} when
the incommensurate
charge $\tilde e$ was a fraction of the electron
charge and the Polyakov loop operator transforms
under a $Z_N$ subgroup of $Z$.
Thus, as in finite temperature Yang-Mills theory, the Abelian
version of the Polyakov loop operator can be used as
an order parameter for the $Z$
symmetry.

The $Z$ symmetry is related to charge screening and
confinement in quantum electrodynamics in the same way as the $Z_N$
symmetry of finite temperature Yang-Mills theory.  The correlators of
Polyakov loop operators measure the free energy of the
electrodynamic system in the presence of
a distribution of static charged sources.
The two-point function, for example,
\begin{equation}
e^{-F_{\tilde e,-\tilde e}(\vec x,\vec y)/T}
=\langle P_{\tilde e}(\vec x)P_{-\tilde e}(\vec y)\rangle
\end{equation}
measures the effective interaction potential between particles
with charges $\tilde e$ and
$-\tilde e$ and positions $\vec x$ and $\vec y$, respectively.
In a de-confined phase, we
would expect Debye screening and the asymptotic
form of the potential at large separations
to decay exponentially with
the Debye mass of the photon.
This would imply that the correlator of two Polyakov
loop operators approaches a constant at large separations.
This implies spontaneous breaking
of the $Z$ symmetry.

In a confined phase, there should be a string tension, and
$F_{\tilde e,-\tilde e}(\vec x,\vec y)$ increases with separation.
 This would give a decay of
the two-point correlator of Polyakov loops
consistent with a $Z$ symmetric phase.

In 3+1-dimensional quantum electrodynamics, we would expect that, at
least in the physically observed weak coupling regime, the $Z$
symmetry is broken spontaneously at all temperatures.  It has recently
been argued~\cite{gss} that in 2+1-dimensional parity invariant
electrodynamics, at least if the electron mass is large enough, both
the confined and de-confined phases should exist with a
Kosterlitz-Thouless type of phase transition between them at some
finite temperature. In the following we shall examine the case of
1+1-dimensional electrodynamics.  There, when the mass of the electron
is non-zero, the dimension of the space is too low to allow a phase
transition.  Based on the results of \cite{cjs} one can expect that
the theory exists in a confined phase at any temperature (except for
some specific $\theta$-vacua found in ~\cite{bsrs}).

On the other hand, we will be able to show that the
$Z$ symmetry is spontaneously broken in 1+1-dimensional electrodynamics
when the electron is massless, i.e. in the Schwinger model.

\subsection{$Z$ symmetry of the Schwinger model}

In this paper we shall examine the expectation value of the Polyakov
loop operator in the Schwinger model.  The $Z$ symmetry transforms the
temporal component of the gauge field as
\be
A_0\rightarrow A_0+2\pi n T/e\ ,\quad \quad n\in Z\ \ .
\label{zsim}
\ee
We shall give an interpretation for the $Z$ symmetry in the Hamiltonian
formalism in terms of the quantization of charge, in states of the thermal
ensemble.  If $Z$ is a good symmetry, all charged states through which
the thermal system fluctuates have charges which are quantized in units of the
electron charge.  If $Z$ is spontaneously broken, there are quantum states
available which have arbitrary charge.  If $Z$ were broken to a subgroup,
$Z_N$, this would imply that there were
fractionally charged states with charges
quantized in units of $e/N$ where $e$ is the charge of the dynamical electron.
An explicit realization of the latter breaking pattern may be of relevance for
applications to one dimensional condensed matter systems.

Our analysis of the finite temperature Schwinger model
with one flavor of fermions leads us to the following results:

\noindent
{\bf i.)}In the one-dimensional Coulomb gas, which can be regarded
as a certain limit of electrodynamics which has very massive charged
particles, the $Z$ symmetry breaking
problem resembles that of the quantum pendulum problem, or 1-dimensional
sine-gordon theory.
The $Z$ symmetry is unbroken at all temperatures,
corresponding to a confining state.

\medskip
\noindent
{\bf ii.)} In the Schwinger model where the space is a circle with
circumference $L$ and with periodic boundary
conditions for both the photon and electron fields, we compute the
expectation value of the Polyakov loop operator and its correlators.
We find that the expectation value of the Polyakov Loop operator with
electric charge $\tilde e$ an integer multiple of the electron charge
is a non-zero computable constant.  When the charge in the loop
operator is not an integer multiple of the electron charge, the
expectation value of the Polyakov loop operator vanishes,
\begin{equation}
\left\langle \exp\left\{ i\tilde e\int_0^{1/T}d\tau A_o(\tau,x)
\right\}\right\rangle=0~~~{\rm if}~\tilde e\neq~{\rm integer}~\cdot e
\nonumber
\end{equation}
at all temperatures $T$.  This can be seen as the consequence of the
discrete symmetry (\ref{zsim}) which is realized in an
unbroken phase when the volume is finite.

\medskip
\noindent
{\bf iii.)} In the infinite volume limit, the $Z$
symmetry is spontaneously broken.  This is seen by
examining the following limits:
\begin{equation}
\lim_{\vert x-y\vert\rightarrow\infty}\left( \lim_{L\rightarrow\infty}
\left\langle  \e{i\tilde e\int_0^{1/T}d\tau A_o(\tau,x)}~\e{-i\tilde
e\int_0^{1/T} d\tau A_o(\tau,y)}\right\rangle\right)= ~{\rm
constant}~\neq0 \label{lrc}
\end{equation}
for any charge $\tilde e$ and at all temperatures.  We interpret this
as implying that the thermal state of the Schwinger model can screen
arbitrary external charges.  The exact form of the correlator is known
and the asymptotic, exponential decay of the correlation function is
governed by the Schwinger mass of the photon, $\ms^2={\rm e}^2/\pi$.

\medskip

\subsection{Symmetry breaking in one dimension?}

The result that the $Z$ symmetry is spontaneously broken was
anticipated by Hansson, Nielsen and Zahed ~\cite{hnz}.  It is
surprising in the sense that, as we shall argue in the following sections,
the effective action for the Polyakov
loop operator is an one dimensional field theory with a discrete
$Z$ symmetry.  Normally such symmetries cannot be spontaneously
broken, as the long-range correlations described by (\ref{lrc}) are
forbidden by the accompanying strong infrared effects.  From another
viewpoint, the ordered state of the broken symmetry theory is unstable
to the condensation of domain walls.

This can be understood by a simple argument: If we consider a one
dimensional system with $N$ sites and $n$ domain walls, the entropy of
the state can be estimated by noting that the domain walls can be
arranged in $ \mat{c}{N\cr n\cr}$ ways, leading to entropy
 \be
 S=\ln\mat{c}{N\cr n\cr}
\ee
 If the domain wall has energy $\epsilon$ the free energy at
temperature $\bar T$ for large N and n is then given by
\begin{equation}
F= n\epsilon+\bar T\Bigl( n\ln n+(N-n)\ln(N-n)-N\ln N\Bigr)
\end{equation}
Note that for all values of the domain wall energy $\epsilon$ and
temperature ${\bar T}$, the entropy always grows faster than the
energy as $n$ is increased.  This leads to a condensation of domain
walls.  The equilibrium number of domain walls has a Fermi-Dirac
distribution
\begin{equation}
\left\langle  n\right\rangle =N\frac{ \e{-\epsilon/\bar T}}{
1+\e{-\epsilon/\bar T}}
\end{equation}
If the size of the system is $Na$ where $a$ is the lattice spacing,
the correlation length is of order the mean distance between domain
walls,
\begin{equation}
\xi\approx (1+\e{\epsilon/\bar T})\ a
\end{equation}
which is always small, of order the ``lattice spacing'' or inverse
ultraviolet cutoff.  Thus, domain wall condensation would seem to
always destroy one-dimensional order.

We shall argue that the Schwinger model evades domain wall
condensation at all finite temperatures by having domain walls with
infinite energy.  This occurs because the domain walls are actually
instantons in a static gauge.  The fermion determinant vanishes on
instanton configurations, giving the instantons an infinite free
energy. Thus, the only way out of the above argument, that
$\epsilon/{\bar T} =\infty$, is actually realized in the Schwinger
model.

When the electrons have a mass, one can expect that the domain wall
energy for small mass diverges logarithmically, $\epsilon\approx-\bar
T\ln(m/\mu)$, for small $m$ where $\mu$ is a dimensional constant
related to the fermion mass $m$ and the confining scale which is given
by the electric charge $e$.  Thus, if the electron in the Schwinger
model had non-zero mass the domain walls would have finite energy, the
correlation length would be
\begin{equation}
\xi\approx (1+\mu/m)\ a
\end{equation}
and the domain wall condensation would ruin the symmetry breaking at all
temperatures, apparently even in the zero temperature limit.

\subsection{Deconfinement versus superconductivity}

There is another interpretation of the physical state of the Schwinger
model alternative to confinement.  The fact that the photon has a mass
can be interpreted as the Schwinger model being a superconductor or,
since in one dimension there is no possibility of magnetic fields and
therefore no Meissner effect, a perfect conductor.  This is seen by
considering the current induced in the Schwinger model ground state by
an external electric field which can be obtained from the exact
identities for current conservation
\begin{equation}
\nabla_\mu \langle j_\mu(x)\rangle_{A}=0
\end{equation}
and the axial anomaly equation which can be presented as
\begin{equation}
\epsilon_{\mu\nu}\nabla_\mu \langle j_\nu\rangle_A=\frac{e^2}{2\pi}
\epsilon_{\mu\nu}\nabla_\mu A_\nu^{\rm ext}
\end{equation}
which makes use of the kinematical identity relating the axial and
vector currents of two dimensional fermions
\begin{equation}
\langle j^5_\mu\rangle_A=i\epsilon_{\mu\nu}\langle j_\nu\rangle_A
\end{equation}
The above equations have the solution
\begin{equation}
\langle
j_\mu(x)\rangle_A=\frac{e^2}{\pi}\frac{\epsilon_{\mu\nu}\nabla_\nu}{-\nabla^2}
E^{\rm ext}
\end{equation}
in terms of the external electric field, $E^{\rm ext}$.  This is
a superconducting response.

For example, if the electric field is spatially constant, it has the
solution
\begin{equation}
\langle J_0\rangle_A=0  ~~,~ \langle J_1\rangle_A=\frac{e^2}{\pi} E^{\rm ext} t
\end{equation}
where the current increases linearly with time.

This superconducting response can lead to a super-screening of
electric fields which would otherwise be caused by external charges.
We propose this as an alternative to the other obvious interpretation
of the breaking of the $Z$ symmetry, the loss of confinement.

Our results regarding the $Z$ symmetry breaking support the
conclusions of Refs. ~\cite{rrs,bsrs,brr}, where the concept of
screening versus confinement in 1+1 dimensional field theories was
discussed.  There, deconfinement was interpreted as arising from
liberation of so called bleached states, with the charges of the
deconfined fermions being completely screened.  The ``bleached
states'' were originally found in the operatorial solution of the
Schwinger model in ~\cite{ls}. In later works ~\cite{cf1,mps} they
were argued to be unphysical in that they are created by non-local
operators and therefore are unattainable by the action of operators
within the algebra of local observables.  The strong infrared effects
driving the theory in infinite volume are essentially non-local,
however, so we shall interpret our results as indicating the possible
emergence of ``bleached states'' in the finite temperature infinite
volume limit.

\medskip

In the next Section, we shall present two simple examples where the
realization of the $Z$ symmetry is in the unbroken phase.  In the
subsequent Section we shall review the Hamiltonian formulation of the
Schwinger model.  It is somewhat independent of the rest of the paper
and is intended mainly to fix notation and remind the reader of the
standard picture.  In Section 4, we describe the path integral
representation of the partition function of the Schwinger model at
finite temperature.  We also introduce the effective action for the
Polyakov loop operator and make explicit the physical interpretation
of the $Z$ symmetry.  In Section 5, we calculate the Polyakov loop
expectation values and prove results ii) and iii).  in Section 6, we
present a discussion of our conclusions.


\section{Two Simple Examples}

Before we solve for the Polyakov loop correlator in the Schwinger
model, let us consider the following examples.

\subsection{Free electrodynamics in 1+1 dimensions}

First, let us consider the case of two
dimensional pure U(1) gauge theory.  The correlator of Polyakov loop
operators is given by
\begin{equation}
\langle  \prod_j e^{ie_j\int_0^{1/T}d\tau A_0(\tau,x_j)}\rangle
\equiv
{\int dA_\mu(x) e^{-\int_0^{1/T}dxd\tau F_{01}^2/2}
\prod_j e^{ie_j\int_0^{1/T}d\tau A_0(\tau,x_j)}   \over
\int DA_\mu(x) e^{-\int_0^{1/T}dxd\tau F_{01}^2/2 } }
\label{101}
\end{equation}
where the finite temperature path integral is done with periodic
boundary conditions.  The partition function has the formal symmetry
\begin{equation}
A_0(x,\tau)\rightarrow A_0(x,\tau)+~{\rm constant}
\end{equation}
which, because of the absence of the dynamical
electron field, is larger than the
$Z$ symmetry.

It is straightforward to perform the gaussian integral in (\ref{101})
to obtain the exponential of the 1-dimensional Coulomb energy.  The
result has an infrared simgularity unless
\begin{equation}
\sum e_j=0
\end{equation}
If this constraint is obeyed, we obtain the expression
\begin{equation}
\langle  \prod_j e^{ie_j\int d\tau A_0(x_j,\tau)}\rangle
=\exp\left(-\sum_{ij}\frac{e_ie_j}{2T}\vert x_i-x_j\vert\right)
\end{equation}
This is the usual confining 1-dimensional coulomb potential.  It
corresponds to a state where the symmetry under translation of $A_0$
in the path integral is unbroken.  This is seen, for example, in the
correlator $\langle  e^{ie\int A_0(\tau,x)} e^{-ie\int
A_0(\tau,y)}\rangle $ which has the asymptotic form
\begin{equation}
\lim_{\vert x-y\vert\rightarrow\infty}~
\langle  e^{ie\int A_0(\tau,x)} e^{-ie\int A_0(\tau,y)}\rangle
=\lim_{\vert x-y\vert\rightarrow\infty}~e^{-e^2\vert x-y\vert}
=0
\end{equation}
The cluster decomposition implies that the symmetry is unbroken at any
finite temperature.

\subsection{1-Dimensional Coulomb gas}

The 1-dimensional Coulomb gas is implicitely solvable through the
Gibbs ensemble calculation of Ref. ~\cite{lenard}.  Here, we obtain
the grand canonical partition function of a neutral Coulomb gas by the
following construction.  Consider the statistical mechanics of a state
with $m$ classical particles with charge $e$ occupying positions
$x_1,\dots x_m$ and $n$ classical particles with charge $-e$ occupying
positions $y_1,\ldots,y_n$.  The partition function is given by
\begin{equation}
\frac{ \int dA_\mu e^{-\int_0^{1/T}F_{\mu\nu}^2/4}~e^{ie\int_0^{1/T}d\tau
(\sum_1^m A_0(x_j)-\sum_1^n A_0(y_j))} }
{ \int dA_\mu e^{-\int_0^{1/T}F_{\mu\nu}^2/4} }
\end{equation}
We multiply by the statistical factor for identical particles,
$1/m!n!$ and a fugacity paramter
$\lambda^{m+n}$, average over positions $x_i$ and $y_i$ by integrating
and sum over $m$ and $n$ to obtain the partition function
\begin{equation}
Z[\lambda,T]=
\frac{ \int dA_\mu e^{-\int_0^{1/T}F_{\mu\nu}^2/4
+\lambda\int dx e^{ie\int_0^{1/T}d\tau A_0(\tau,x) + {\rm c.c.} } } }
{ \int dA_\mu e^{-\int_0^{1/T}F_{\mu\nu}^2/4} }
\end{equation}

If we fix the gauge
\begin{equation}
\frac{\partial}{\partial\tau}A_0(\tau,x)=0
\end{equation}
we can do the integral over $A_1$ and obtain the one-dimensional
sine-gordon theory
\begin{equation}
Z[\lambda,T]=\int \prod_x dA_0(x)\exp\left(
-\int dx\left(\frac{T}{2} (dA_0(x)/dx)^2-\lambda\cos (eA_0(x)/T) \right)\right)
\label{sg}
\end{equation}

The effective action for $A_0(x)$ explicitly has the symmetry under
the shift $A_0\rightarrow A_0+2\pi n T/e$ In the one-dimensional
system (\ref{sg}) this symmetry cannot be spontaneously broken for any
values (aside from zero or infinity) of the parameters $\lambda$ and
$T$. Thus, the expectation value of the Polyakov loop must vanish
unless it has charge $e$.  This we interpret as confinement.  There is
no confinement-deconfinement transition in this model.


\section{Hamiltonian Formalism}

\subsection{Hamiltonian, gauge constraints and theta-states}

We shall consider 1+1-dimensional electrodynamics defined on a compact
space, $x\in[0,L]$.  We begin by working in the canonical, Hamiltonian
formalism.  The Hamiltonian is derived from the action
\begin{equation}
S=\int dt\int_0^L dx\left( \half F_{01}^2- \bar\psi
\gamma\cdot(i\nabla+e A)\psi\right)
\label{action}
\end{equation}
In this action, the canonical momentum conjugate to the spatial
component of the gauge field ($A_1(t,x)$, which we shall shortly
rename $A(t,x)$) is the electric field $E(t,x)\equiv
F_{01}(t,x)=\nabla_t A_1-\nabla_x A_o$.  The momentum conjugate to the
fermion $\psi(t,x) $ is $i\psi^{\dagger}(t,x)$.  The non-vanishing
equal time (anti-) commutation relations are therefore
\begin{eqnarray}
\left[ A(x), E(y)\right]&=&i\delta(x-y)\nonumber \\
\left\{ \psi (x),\psi^{\dagger}(y)\right\}&=&\delta(x-y)
\label{com}
\end{eqnarray}
The temporal component of the gauge field, $A_o(t,x)$, appears in the
action (\ref{action}) without time derivatives and acts as a Lagrange
multiplier to impose the constraint of invariance under gauge
transformations.  The Hamiltonian is obtained as
\begin{equation}
H=\int_0^L dx\left( {1\over2}E^2(x)+ \psi^{\dagger}(x)\alpha(i\nabla
+e A(x))\psi (x)\right)
\label{ham}
\end{equation}
where $\alpha=\gamma_5=\gamma_o\gamma_1$ is a $2\times2$ Hermitean
Dirac matrix and $\nabla\equiv d/dx$.  The massless Dirac Hamiltonian
can be decomposed into Hamiltonians for left and right movers as
\begin{equation}
H_{\rm Dirac}=\int_0^L dx\left(\psi^{\dagger}_L(x)(i\nabla+eA(x))\psi_L(x)
-\psi_R^{\dagger}(x)(i\nabla+eA(x))\psi_R(x)\right)
\end{equation}

All fields have periodic boundary conditions in space,
\begin{eqnarray}
A(L)&=&A(0)~~~~~~~E(L)\ =\ E(0)
\nonumber\\
\psi (L)&=&\psi (0)~~~~~~\psi^{\dagger}(L)\ =\ \psi^{\dagger}(0)
\label{pbc}
\end{eqnarray}
The Hamiltonian and commutation relations must be supplemented by the
first class constraint, or Gauss' law, which is the operator
obtained by taking a functional derivative of the action
(\ref{action}) by $A_o$,
\begin{equation}
G(x)\equiv -\nabla E(x)-e \psi^{\dagger}(x)\psi (x)\sim0
\label{gauss}
\end{equation}
and ensuring that the quantum states are invariant under
time-independent gauge transformations.  The latter are generated by
the operator
\begin{equation}
G[\chi]\equiv \int_0^L dx \chi(x)G(x) \label{gee}
\end{equation}
where $\chi(x)$ is a periodic function, $\chi(L)=\chi(0)$. The action
of the operator \refe{gee} is
\begin{eqnarray}
\e{iG[\chi]}A(x)\e{-iG[\chi]}&=&A(x)+\nabla\chi(x)\nonumber \\
\e{iG[\chi]}E(x)\e{-iG[\chi]}&=&E(x)\nonumber \\
\e{iG[\chi]}\psi (x)\e{-iG[\chi]}&=&\e{ie\chi(x)}\psi (x)\nonumber \\
\e{iG[\chi]}\psi^{\dagger}(x)\e{-iG[\chi]}&=&\e{-ie\chi(x)}\psi^{\dagger}(x).
\label{gauge}
\end{eqnarray}
 This is a symmetry of the Hamiltonian (\ref{ham}) and of the
commutation relations (\ref{com}), which preserves the boundary
conditions \refe{pbc}.

There is a larger class of gauge transformations under which the
Hamiltonian and commutation relations are invariant, which preserve
the boundary conditions (\ref{pbc}) and which are not generated by the
Gauss operator $G[\chi]$.  These have gauge functions which are not
strictly periodic, but obey the condition
\begin{equation}
\chi_n(x+L)=\chi_n(x)+2\pi n/e \ .
\label{lggauge}
\end{equation}
This guarantees that both the electron operator and the gauge field
boundary condition is unchanged. Such
`large' gauge transformations can always be expressed as a periodic
gauge transformation plus a representative of the large, non-periodic
transformations as
\begin{equation}
\chi_n(x)=\chi_o(x)+ 2\pi nx/Le \ ,
\label{decomp}
\end{equation}
where $\chi_o(x)$ is periodic.

Large gauge transformations are implemented by an unitary operator
$\exp(iG_{\ell}[\chi_n])$, where the Hermitean operator is
\begin{equation}
G_{\ell}[\chi_n]= \int_0^L\left( \nabla\chi_n (x)E(x)-e\chi_n(x)
\psi^{\dagger}(x)\psi (x)\right) \label{lololo}
\end{equation}
Using (\ref{decomp}), this generator can be written as a large gauge
transformation generator and a Gauss' operator
\begin{equation}
G_{\ell}[\chi_n]=G[\chi_o]+\frac{2\pi n}{Le}\int_0^L dx E(x)-
\frac{2\pi n}{L}\int_0^L dx x \psi^{\dagger}(x)\psi (x)
\label{lgt}
\end{equation}

\bigskip

The quantization of the model with commutator algebra (\ref{com}),
Hamiltonian ({\ref{ham}) and constraint (\ref{gauss}) can proceed
in two different ways.  First, one can solve the constraint
(\ref{gauss}) at the classical level by imposing an auxiliary
condition on the remaining degrees of freedom.  The second and
equivalent approach, which we shall pursue in the following, is to
quantize the dynamical system specified by (\ref{ham}) and (\ref{com})
as it is.  Then, on the Hilbert space which represents the algebra
(\ref{com}) and where the Hamiltonian is diagonalizable, we shall
impose the physical state condition
\begin{equation}
G[\chi_o]\ \vert{\rm physical~state}>\ =\ 0
\end{equation}
The physical states are thus invariant under all periodic gauge
transformations.

However, they need not be invariant under the set of all gauge
transformations.  In fact, it is only necessary that they transform
under a unitary irreducible representation of the coset of
time-independent gauge transformations modulo the periodic ones.  The
coset group is isomorphic to the translation group of the integers,
$Z$, whose unitary irreducible representations are one dimensional
phases, $\e{i\theta n}$.  Thus, if we implement a large gauge
transformation using the operator $G_{\ell}[\chi_n]$, the physical
states should transform as
\begin{equation}
\e{iG_{\ell}[\chi_n]}\ \vert{\rm
physical~state},\theta>\ =\ \e{in\theta}\ \vert{\rm physical~state},\theta>
\end{equation}
In this way, the physical states are characterized by a theta-angle.

Like the theta-angle of non-Abelian gauge theories in four spacetime
dimensions ~\cite{cdg,jr}, there exists a canonical transformation
which removes the theta-angle from the states and introduces a theta
term in the action.  The unitary operator which implements this
transformation is
\begin{equation}
U(\theta) =\exp\left( -i \theta {e}/{2\pi}~\int_0^L dx A(x)\right)
\end{equation}
In the new system, the theta angle is absent from the physical states and
the electric field operator is altered.  The new hamiltonian is
\begin{equation}
H=\int_0^L dx\left( \half (E(x)+ \theta {e}/{2\pi})^2+
\psi^{\dagger}(x) \alpha(i\nabla+e A)\psi (x)\right)
\end{equation}
In this way, one sees that theta has the interpretation of a constant
background electric field, as discussed by Coleman, Jackiw and
Susskind ~\cite{cjs}.  In the original action, taking into account the
role of $A_o$, (\ref{action}) is modified by the addition of the
conventional $\theta$ term,
\begin{equation}
S=\int d^2x\left( \half F_{01}^2 +\theta F_{01}- \bar\psi
\gamma\cdot(i \nabla+ e A)\psi (x)\right)
\end{equation}

It turns out that, in the massless Schwinger model, the physical
states do not depend on $\theta$.  It is also possible to see that the
parameter $\theta$, which appears with the topological term, in the
action is invariant.  In the following we shall retain the
theta-dependence in order to demonstrate the theta-independence of the
partition function.

In the next Section we shall discuss the construction of the path
integral representation of the thermodynamic partition function.


\section{Path integral representation of the partition function}

The thermodynamic description of field theory is most conveniently
obtained from the partition function which for a gauge theory is
gotten by taking a trace over physical states of the Gibbs
distribution, $\e{-H/T}$, where $T$ is the temperature and we work in
units where the Boltzmann's constant as well as the Planck's constant
and the speed of light are equal to one.  In constructing the
partition function it is convenient to consider all the states which
represent the commutator algebra (\ref{com}) and insert a projection
operator which projects over the physical states, and onto a sector
with a fixed vacuum angle $\theta$.  The trace is thus given using a
complete set of states which represent (\ref{com}),
\begin{equation}
Z[T]=\sum_s <s\vert \e{-H/T}P_\theta \vert s>
 \label{54}
\end{equation}
The appropriate projection operator is constructed from the unitary
operator which generates gauge transformations
\begin{equation}
P_\theta =\frac{1}{\rm Vol.G}\sum_n \e{-in\theta}\int [d\chi_n] \e{iG[\chi_n]}
\end{equation}
where we have integrated over all time independent gauge
transformations and divided by the (infinite) volume of the gauge
group.  This results in the expression for the partition function
\begin{equation}
Z[T]= \frac{1}{\rm Vol.G}\sum_n\e{-in\theta}\int[d\chi_n]
 \e{-S_{\rm eff}[\chi_n]}
\label{part}
\end{equation}
where the effective action for the gauge degrees of freedom is given
by the twisted trace
\begin{equation}
\e{-S_{\rm eff}[\chi]}=\sum_s <s\vert \e{-H/T}\vert s^{\chi_n}>
\label{effec}
\end{equation}

This effective action has a standard path integral representation;
in phase space it is given by
\begin{equation}
\e{-S_{\rm eff}[\chi]}=\int
\prod_{x\in[0,L]}\prod_{\tau\in[0,1/T]}d\psi(\tau,x)
d\psi^{\dagger}(\tau,x) dA(\tau,x)
dE(\tau,x)~\e{-S_E[\psi,\psi^{\dagger},A,E]} \label{efac}
\end{equation}
where the Euclidean first order action is
\begin{equation}
S_E=\int_0^{1/T}d\tau\int_0^L dx\left(iE\dot A+\half E^2-
\psi^{\dagger}\left[\nabla_\tau+i\alpha\nabla_x+e\alpha
A \right]\psi \right).
\end{equation}
The electric field $E(\tau,x)$ has open boundary conditions in
time, and the other integration variables have twisted (anti-)
periodic boundary conditions,
\begin{eqnarray}
A(1/T,x)&=&A(0,x)-\nabla\chi_n(x) \nonumber\\
\psi (1/T,x)&=&-\e{ie\chi_n(x)}\psi (0,x) \nonumber\\
\psi^{\dagger}(1/T,x)&=&\psi^{\dagger}(0,x)\e{-ie\chi_n(x)}
\end{eqnarray}
The Gaussian integral over the canonical momentum $E(x)$ is
performed to yield, up to a temperature and volume dependent but
$\chi_n$-independent overall factor,
\begin{equation}
\e{-S_{\rm eff}[\chi_n]}=\int
\prod_{x\in[0,L]}\prod_{\tau\in[0,1/T]}d\psi(\tau,x)
d\psi^{\dagger}(\tau,x)dA(\tau,x)~\e{-S_E[\psi,\psi^{\dagger},A]}
\label{pi}
\end{equation}
where now the Euclidean action is given by
\begin{equation}
S_E=\int_0^{1/T}d\tau\int_0^L dx\left(\half (\dot
A)^2-\psi^{\dagger}\left[\nabla_\tau+
i\alpha\nabla_x+e\alpha A \right]\psi \right)
\end{equation}

The boundary conditions can be untwisted by a non-periodic gauge
transformation.  This is what normally restores $A_o$, the temporal
component of the gauge field to the Euclidean path integral.
A suitable
non-periodic gauge transformation
redefines the integration variables as
\begin{eqnarray}
 A(\tau,x) &\mapsto& A(\tau,x)-\nabla \chi_n(x)T\tau \nonumber\\
 \psi (\tau,x)&\mapsto& \e{-ie\chi_n(x)T\tau}\psi(\tau,x) \\
 \psi^{\dagger}(\tau,x) &\mapsto& \psi^\dagger(\tau,x)\mm
\e{ie\chi_n(x)T\tau} \nonumber
\end{eqnarray}
 Note that the spatial boundary conditions for
the fermi fields are now changed.
The resulting path integral now has the action
\begin{equation}
S_E=\int_0^{1/T}d\tau\int_0^L dx\left(\half(\dot
A)^2+\half T^2(\nabla\chi_n)^2-\psi^{\dagger}\left[\nabla_\tau
-ieT\chi_n+ i\alpha\nabla_x+e\alpha A \right]\psi \right)
\label{act}
\end{equation}
 and the boundary conditions are
\begin{eqnarray}
\psi(1/T,x) \ =\ -\psi(0,x)
	~~~~~~~\psi (\tau,L) &=& \e{2\pi i n T\tau}\psi (\tau,0) \nonumber\\
\psi^{\dagger}(1/T,x)\  =\  -\psi^{\dagger}(0,x)
{}~~~~~\psi^{\dagger}(\tau,L) &=& \e{-2\pi i n
T\tau}\psi^{\dagger}(\tau,0)
	\label{bc}\\
A(1/T,x) \ =\ A(0,x) ~~~~~~~~~ A(\tau,L)&=& A(\tau,0) \nonumber\\
		\chi_n(L)&=&\chi_n(0)+2\pi n/e  \nonumber
\end{eqnarray}

In order to compute the effective action for $\chi_n(x)$ we must
compute the path integral (\ref{efac}) with the Euclidean action
(\ref{act}) and the boundary conditions (\ref{bc}).

The effective electromagnetic field tensor is given by
\begin{equation}
F_{01}(\tau,x)=\dot A(\tau,x)-\nabla A_o(\tau,x)
\end{equation}
where we identify the temporal component of the gauge field in a
static gauge ($\nabla_\tau A_o=0$) as
\begin{equation}
A_o(x)\equiv T\chi_n(x) \label{static}
\end{equation}
This field configuration has instanton number $n$, as it can be seen from
the integral
\begin{equation}
-\frac{e}{2\pi}\int_0^{1/T}d\tau\int_0^L dx F_{01}(\tau,x)
=n
\end{equation}
where we have made use of the fact that the field $A(\tau,x)$ has
periodic boundary conditions in both $\tau$ and $x$.  Thus, the
effective vector potential fields in the $n$'th sector are just those
which are $n$ instantons in a static gauge.

\subsection{$Z$ symmetry} \label{symsect}

We consider the following change of integration variable in the path
integral (\ref{pi}, \ref{act}):
\begin{eqnarray}
\psi(\tau,x) & \mapsto & \e{2\pi i kT\tau}\psi(\tau,x)
\nonumber\\
\psi^{\dagger}(\tau,x) & \mapsto & \e{-2\pi i
kT\tau}\psi^{\dagger}(\tau,x) \ .
\end{eqnarray}
When $k$ is an integer, the boundary conditions (\ref{bc}) are
unchanged by this substitution and the Jacobian in the path integral
measure is one.  The net effect is to replace the variable $\chi_n(x)$
by $\chi_n(x)+2\pi k/e$.  Thus, the effective action for $\chi_n$ has
the symmetry\footnote{Note that this could in principle be only a
formal symmetry of the path integral.  Here, it survives path
integration because of anomaly cancellation, similar to the
cancellation of gauge anomalies.}
\begin{equation}
S_{\rm eff}[\chi_n]=S_{\rm eff}[\chi_n +2\pi k/e]
\label{sym}
\end{equation}
This is a large gauge symmetry analogous to (\ref{lggauge}) which is
associated with the periodic nature of the space in the temporal
rather than spatial direction.  However, being associated with
Euclidean time, it cannot be a basic symmetry of the theory, it is
rather an effective symmetry of the Euclidean path integral.  We shall
presently discuss its interpretation in the Hamiltonian formalism.

In order to obtain a physical interpretation of this symmetry, we
consider a modification of electrodynamics where there is an array of
static external charges $e_1,\ldots,e_k$ located at positions
$x_1,\ldots,x_k$.  This can be taken into account by a modification of the
Gauss' law to
\begin{equation}
-\nabla E(x)-e  \psi^{\dagger}(x)\psi (x)-\sum_{j=1}^k
e_j\delta(x-x_j)\sim0
\end{equation}
The sole effect of this modification in the partition function
(\ref{part}) is the replacement
\begin{equation}
\e{-S_{\rm eff}[\chi_n]}\rightarrow \e{-S_{\rm eff}[\chi_n]}\prod_{j=1}^k
\e{-ie_j\chi_n(x_j)}
\label{trans}
\end{equation}
This implies that the k-point correlator
\begin{equation}
\left\langle \prod_{j=1}^k \e{-ie_j\chi(x_j)}\right\rangle\equiv{
\sum_n \e{-in\theta}\int [d\chi_n] \e{-S_{\rm
eff}[\chi_n]}\prod_{j=1}^k \e{-ie_j\chi_n(x_j)} \over \sum_n
\e{-in\theta}\int[d\chi_n]\e{-S_{\rm eff}[\chi_n] } } \label{expect}
\end{equation}
is the ratio of the partition function for the electrodynamic system
in the presence of the external charges to the partition function of the
same system in the absence of external charges.  Thus, the free energy
of the system with charges, compared to that without is given by
\begin{equation}
F(e_j,x_j)-F_o
=-T\ln\left\langle \prod_{j=1}^k \e{-ie_j\chi(x_j)}\right\rangle
\label{fe}
\end{equation}
where the bracket $\langle ~~\rangle$ is the average over the fields
$\chi_n$ with the measure $\sum_n \exp(-S_{\rm
eff}[\chi_n]+in\theta)$.  Thus, the correlators of the exponential
operators measure the Coulomb energy of external charges. In this way
they probe the ability of the system to screen external charges.

The symmetry of the effective field theory under the Z transformation,
if it is not
spontaneously broken, poses a restriction on the correlators which can
be non-zero -- and therefore it restricts which arrays of external
charges can have finite free energy (\ref{fe}).  In finite volume,
this symmetry is certainly realized canonically and the result is that
any expectation value of the form (\ref{expect}) averages to zero when
the charges do not add to multiples of the electron charge,
\begin{equation}
\sum_i e_i=me  \ .
\nonumber
\end{equation}
Whether this symmetry persists in the infinite volume limit
$L\rightarrow\infty$, is an interesting question which we shall discuss
in following sections.

{}From the definition \refe{lololo} of the generators of large gauge
transformations, we see that the $Z$ transformation \refe{sym}
changes the generators by
\be
  G_l[\chi_n] \mapsto G_l[\chi_n]  - 2\pi k\int_0^L dx\ \psi^\dagger\psi\ .
\ee
 Accordingly, going back to the definition of the partition function in
(\ref{54}), we see that the $Z$ transformation corresponds
to the replacement of the density matrix $\e{-H/T}$ by the operator
 \be
 \e{-H/T}\exp\left(-2\pi i k\int_0^L dx\ \psi^{\dagger}\psi\right)
 \label{ferminum}
\ee
 in the trace over the physical states. Since all of the physical
states obey Gauss' law, with finite $L$ and periodic boundary
conditions, they have zero fermion number. Therefore, the exponential
of the fermion number is trivially the unit matrix on the space of
physical states.  The question arises whether this fact persists in
the infinite volume limit, or if there are states with arbitrary
fermion number.

\subsection{Computing the Fermion Determinant}

Now we want to calculate the effective action for $\chi$, the
time-component of the gauge field in a static gauge \refe{static}.
For this we first integrate out the fermions from the path integral
(\ref{efac},\ref{act}).

When the fermion mass is zero, the Dirac operator has zero modes for
any of the field configurations with $n\neq0$.  Thus, all the terms in the
sum over $n$ except the term with $n=0$ vanish.

The Dirac operator has zero modes as a consequence of the
Atiyah-Singer index theorem (see e.g. Ref. \cite{egh}).
An explicit demonstration in the Scwhinger model  can be found
in Sachs and Wipf ~\cite{sw}. The Dirac operator has exactly $\abs{n}$
zero modes  in the $n$-instanton sector.

For expectation values of Polyakov lines, we need to take into account
only the zero instanton sector, as there are no fermions in our
correlators to soak up the zero modes.  The fermion determinant in the
zero instanton sector is\footnote{Note that compared to the
Minkowskian $\gamma$-matrices used in \refe{action}, the $\gamma$:s
used henceforth obey Clifford relations with an Euclidean metric, thus
absorbing the extra $i$ in Action \refe{act}.}
\begin{equation}
\int d\bar\psi d\psi \e{\int \bar\psi \gamma\cdot(\nabla-ieA)\psi}=
\det\left( \gamma\cdot\nabla -ie\gamma\cdot A\right)
\end{equation}
We begin with a Hodge decomposition of the gauge field
\begin{equation}
A_\mu= \nabla_\mu\phi+\epsilon_{\mu\nu}\nabla_\nu\Omega+
	{2\pi} h_\mu /e
\end{equation}
where $h_\mu$ are the harmonic modes,
\begin{equation}
h_\mu=\frac{e}{2\pi}\frac{T}{L}\int_0^{1/T}d\tau\int_0^L dx
	A_\mu(\tau,x)\ .
\end{equation}
As the gauge field lies in the zero instanton sector, the pure gauge
($\phi$) and coexact ($\Omega$) part have to be strictly periodic
in space, c.f. \refe{bc}.

Using this decomposition, the fermion determinant can be written as
\begin{eqnarray}
\det\left(
\gamma\cdot(\nabla-ieA)\right)&=&\det\left( \e{i\phi+\gamma_5\Omega}
	\mm\gamma\cdot(\nabla-2\pi ih)\mm\e{-i\phi
	+\gamma_5\Omega}\right)\nonumber\\
&=&\det\left( \gamma\cdot(\nabla-2\pi ih)\right)\det \e{2\gamma_5\Omega}
\end{eqnarray}
Here, we have assumed that the determinant of $\e{i\phi}$ is the
inverse of the determinant of $\e{-i\phi}$. This can be shown to be
true using a gauge invariant regularization, e.g. Pauli-Villars
regularization.  The other factor is
\begin{equation}
\det\left(
\e{2\gamma_5\Omega}\right)=\exp\left({\rm tr}2\gamma_5\Omega\right)
\end{equation}
This is a standard computation; the coexact part of the gauge field
carries the chiral anomaly, which can be integrated using any gauge
invariant regularization. Noting that $\triangle\Omega = F_{01}$, we get
\begin{eqnarray}
\det\left( \e{2\gamma_5\Omega}\right)&=&\exp\left(\frac{e^2}{\pi}\int
  	d^2x\mm \Omega(x)F_{01}(x)\right) \nonumber\\
&=&\exp\left(-\frac{e^2}{2\pi}\int d^2x
	F_{01}\frac{1}{-\nabla^2}F_{01}\right) \label{coedet}
\end{eqnarray}
In the field strength $F_{01}$ there is no contribution from the harmonic
modes of $A_\mu$.

The part of the determinant of the Dirac operator containing only the harmonic
components of the gauge field has the form
\begin{equation}
\det\left(\gamma\cdot(\nabla-i2\pi h)\right)=\prod_{m,n}\left[ \left(
(2n+1)\pi T + 2\pi h_o\right)^2+ \left( 2\pi m/L+2\pi h_1\right)^2\right]
\label{hardet}
\end{equation}
which is the square modulus of the chiral determinant
\begin{equation}
\det D_+\equiv\prod_{mn}\left( (2n+1)\pi T+2\pi h_o+2\pi im/L
	+2\pi ih_1\right)
\label{diplus}
\end{equation}
The latter has the $Z$-symmetry invariance $h_0\rightarrow h_0 + k T$,
$k\in Z$, and depends
only on the complex coordinate $u=h_0+ih_1$. One can compute the
determinant \refe{diplus}
by means of a regularization that preserves the $Z$-symmetry but
breaks the holomorphic
factorization of (\ref{hardet}) \cite{algmova}. Namely, one shall
obtain for \refe{diplus} a
$Z$-invariant result that will depend also on the coordinate $\bar u$.
Such a result is the well-known expression of \refe{diplus} in
terms of Jacobi theta and
Dedekind eta functions, \bea
\det D_+ &=& \frac{1}{\eta(i\t)}\mm
   \e{-\pi\t \tilde h_o^2+2\pi i\tilde h_o(1/2 -\tilde h_1)}
	\mm\Theta(\half-\tilh_1+ i\t \tilh_o,\mm i\t) \cr
 	& \equiv & \frac{1}{\eta(i\t)}\ \Theta\Bigl[\begin{array}{cc}
      {\tilh_o}\\{\half-\tilh_1} \end{array}
\Bigr](0,i\t) \label{chidet}
\eea
where the modular parameter is $i\t=iLT$, and the harmonic modes are
rescaled with the corresponding length scales to get dimensionless
quantities:
\be
 \tilh_o = h_o/T\; \tilh_1 = L h_1 \ .
\ee
 For the theta functions, we follow the labelling conventions of
Mumford \cite{mum},
\bea
\Theta(z,i\t)&=&\sum_{n\in{\cal Z}} \exp\left(-\pi\t n^2+2\pi
	inz\right)\\
\Theta\Bigl[\begin{array}{cc}{a}\\{b} \end{array}
\Bigr](z,i\t) &=&
 \e{-\pi\t a^2 + 2\pi ia(z+b)}\Theta(z+b+ i\t a,\mm i\t) \ .
\eea
As announced the $\tilde h_0^2=(u+\bar u)^2/4$ term in (\ref{chidet}) breaks
the
holomorphic factorization of the determinant (\ref{hardet}).
This is a finite local
counterterm that can be added to the effective action in
order to obtain a gauge invariant result~\cite{algmova}.
As a matter of fact, the gauge
symmetry on the  harmonic component $h_0$ of the gauge field is nothing
but the $Z$-symmetry.
Alternatively, one can maintain the holomorphic factorization
of the determinant \refe{hardet}
and loose the $Z$-symmetry invariance~\cite{dhopho}. Our choice is to keep the
$Z$-symmetry invariance.

The total fermion determinant is obtained by combining the coexact
piece \refe{coedet} with the harmonic piece, which is the modulus
square of the chiral determinant \refe{chidet}. Expressing the field
strength in terms of the spatial gauge field $A$ and the static
temporal field $\chi$, we finally get for the effective action
\begin{eqnarray}
\e{-S_{\rm eff}[\chi]}&=&\int [dA] \det( i\gamma\cdot D)
{}~\e{-{\scriptstyle\frac{1}{2}}
  \int dx d\tau \left\{\dot A^2 + T^2 \nabla\chi^2 \right\}} \cr
 &=& \int d\tilh_1 \abs{\eta(i\t)^{-1}\ \Theta\Bigl[\begin{array}{cc}
      { \tilh_o}\\{ \half -\tilh_1} \end{array}
	\Bigr](0,i\t)   }^2~\times \\
 &&\times  \int [d\hat A]\mm \exp\left\{\half \int dx d\tau
\left\{ {\hat A} \left( 1 - \frac{\ms^2}{\nabla^2}
\right) \nabla_\tau^2 {\hat A}
 - T^2 \left( (\nabla_x\hat\chi)^2 + \ms^2\hat\chi^2 \right)
\right\}\right\} \nonumber
\end{eqnarray}
 The hat on the fields means that the harmonic part is removed. We
have denoted the Scwhinger mass by $\ms^2 = {e^2}/{\pi}$.

The gauge choice (\ref{static}) has not fixed completely the gauge,
 since the harmonic part of
the gauge field is unaffected by \refe{static}.
Consequently the $Z$-symmetry is still
present as a residual gauge invariance.

The $\tilh_1$ and $\hat A$ integrations can now be done, yielding (up
to normalization)
\bea
 \e{-S_{\rm eff}[\chi]} &=& \sum_{n\in\Z} \e{-2\pi\t(n+\tilh_o)^2}\mm
\times \e{-{\scriptscriptstyle {T}/{2}} \int_0^L dx \left\{
(\nabla\hat\chi)^2 + \ms^2\hat\chi^2 \right\}} \cr
 &=& \sum_{n\in\Z} \exp\left\{-{T}/{2} \int_0^L dx\left\{
(\nabla\chi)^2 + \ms^2\left(\chi + {2\pi} n/e\right)^2
\right\}\right\}
 \label{finally}
\eea
In order to get a finite result, one has to gauge fix the residual
(spatial) large gauge transformations by restricting the integration
over $\tilh_1$ to the period $[0,1]$ of the integrand.

\eq\refe{finally} provides the form of the effective action
which explicitly realizes the symmetry \refe{sym}.


\section{Polyakov loop correlators}

Now we are in position to calculate expectation values and correlators
of Polyakov loops, and accordingly to decide, whether the $Z$
symmetry \refe{sym} is spontaneously broken.  As indicated in Section
\ref{symsect}, inserting Polyakov loops probes the response of the
theory to static external charges.  We take the external charges to
have charge $\tilde e$.

Due to the zero modes of the Dirac operator, only the zero instanton
sector contributes to the Polyakov loop correlators \refe{expect}.
The expectation value divides in a global and local part:
\bea
 \langle  \e{i \tilde e \chi(x)} \rangle &=& 1/\N \int [d\chi]
\e{-S_{\rm eff}[\chi]}   \e{i \tilde e \chi(x)} \\
&=& \int  d\tilh_o \sum_{n\in\Z} \e{-2\pi \t (n+\tilh_o)^2
 + 2\pi i  \tilh_o {\tilde e}/{e}}
\times \int [d\hat\chi]
 \e{-{\scriptstyle{T}/{2}}\int_0^L dy
\left\{
(\nabla\hat\chi)^2 + \ms^2\hat\chi^2 \right\}}
\e{i\tilde e\hat\chi(x)} \nonumber
\eea
 The global part is easily calculated by Poisson
resummation:\footnote{ Note that as a gauge fixing of large
spatial gauge transformations, the domain of the integration has to be
restricted to the period of the integrand, which depends on the value
of $\tilde e/e$.}
\bea
\int  d\tilh_o \sum_{n\in\Z} \e{-2\pi\t (n+\tilh_o)^2
  + 2\pi i \tilh_o {\tilde e}/{e}}
   &=&
\sqrt{\frac{1}{2\t}}\int  d\tilh_o \sum_{\nu\in\Z}
\e{-{{\pi}/{2\t}}\mm \nu^2 + 2\pi i \tilh_o \left(
{{\tilde e}/{e}} + \nu\right)} \cr
 &=&  \cases{
    \sqrt{\frac{1}{2\t}}\e{-{{\pi}/{2\t}}\mm
({{\tilde e}/{e}})^2},& {\rm if}\ {\tilde e}/{e} \in\Z \\
	0\mm,& {\rm otherwise}
	     }
 \label{poisres}
\eea
The local part can be expressed as
\bea
\int [d\hat\chi] \e{i\tilde e \chi(x)} \e{-{T}/{2}\int_0^L
dy\left\{\hat\chi(y) (\ms^2 - \nabla^2)\hat\chi(y)\right\}} &\sim&
\e{-\frac{\tilde e^2}{2T} K(x,x)}
 \label{localex} \\
 &=& \exp\left\{\frac{\tilde e^2}{2TL\ms^2}- \frac{\tilde e^2}{4T\ms}
\coth \half L\ms \right\}  \nonumber
 \eea
where we used the harmonic oscillator Green's function on the circle
(with the contribution of the zero-modes subtracted),
\be
 K(x,y) = \frac{1}{L}\sum_{n\neq 0}\frac{\e{2\pi
in(x-y)/L}}{\ms^2+4\pi^2n^2/L^2}
 = \frac{1}{2\ms} \frac{\cosh\left( \half L\ms
 (1-2\abs{x-y}/L)\right)}{\sinh \half L\ms }
 \mm-\mm \frac{1}{L\ms^2}\ .
 \label{HOGReen}
\ee
 Combining \refe{poisres} and \refe{localex}, we get for the
Polyakov loop expectation value
\be
 \langle  \e{i \tilde e \chi(x)} \rangle = \cases{
 \exp\left\{- \frac{\tilde e^2}{4T\ms}\coth \half L\ms \right\}\mm,
 & {\rm if}\ {\tilde e}/{e} \in\Z \\
	0\mm,& {\rm otherwise} \label{polexp}
}\ee

This is consistent with the expected unbroken realization of the
$Z$ symmetry for finite volume.  The heat bath screens only external
charges that are integer multiples of the electron charge, by bounding
dynamical fermions from the heat bath with the external charge. This
can be viewed as a proof of the confining nature of finite volume one
dimensional electrodynamics.

In infinite volume, one would expect the contribution of the harmonic
modes to be trivial. Accordingly, one would expect the nonvanishing
value of the Polyakov loop in \eq\refe{polexp} to be valid
irrespective of the value of $\tilde e$, which would be a signal of
$Z$ symmetry breaking.

In order properly to investigate the possible symmetry breaking in the
infinite volume limit, we need the Polyakov-anti-Polyakov loop
correlator.  This is again readily calculated using the effective
action \refe{finally}. The harmonic contributions cancel between the
Polyakov and anti-Polyakov loops, and we are left with the integration
\bea
\left\langle \e{i\tilde e\chi(x)} \e{-i\tilde e\chi(y)} \right\rangle
 &=& 1/\N \int[d\hat\chi] \e{\int_0^L\left\{i\tilde e
  \hat\chi(x')\left(\delta(x'-x) - \delta(x'-y)\right) -
 {T}/{2}\mm\hat\chi(x')(\ms^2 -\nabla^2)\hat\chi(x')\right\}} \cr
 &=& \exp\left\{-{\tilde e^2}/{T}\mm\left(K(x,x)-K(x,y)\right)\right\} \\
&=& \exp\left\{\frac{ -\tilde e^2}{2T\ms}\left(\coth \half L\ms  -
\frac{\cosh\left( \half L\ms  (1-2\abs{x-y}/L)\right)}{\sinh \half L\ms }
 \right) \right\} \nonumber
\eea
In the infinite volume limit we get
\be
\left\langle \e{i\tilde e\chi(x)} \e{-i\tilde e\chi(y)} \right\rangle
\stack{L\to\infty}{\longrightarrow}
\exp\left\{-\frac{\tilde e^2}{2T\ms}\left(1 - \e{-\ms \abs{x-y}}
 \right) \right\} \ . \label{infivol}
 \ee
 This result shows that the system does not cluster decompose:
\be
 \left\langle \e{i\tilde e\chi(x)} \e{-i\tilde e\chi(y)}
 \right\rangle_{L=\infty}
\stack{\abs{x-y}\to\infty}{\longrightarrow}
\exp\left\{-\frac{\tilde e^2}{2T\ms}\right\}
\ee
 This is consistent with the value of the infinite volume Polyakov
loop expectation value anticipated from \refe{polexp}. For $\tilde e
= e$, our results (\ref{polexp}, \ref{infivol})  agree with the
results of \cite{hethos} on Wilson loop expectation values of the zero
temperature Scwinger model on a circle, upon a Wick-rotation.

Equation \refe{infivol} shows that, in the infinite volume limit,
there is off diagonal long-range order, and the $Z$ symmetry
\refe{sym} is thus spontaneously broken. The thermal state of
the Schwinger model can screen arbitrary external charges, and it is
in a deconfined phase.

We interpret our results as indicating the possible presence of the
``bleached'' states of Ref. \cite{ls} in the spectrum of the Schwinger
model. These are states where the charges of the deconfined
fermions are completely screened by the thermal state. Our results
thus support the  screening vs. confinement discussion of
Refs. ~\cite{rrs,bsrs,brr}.

In \eq \refe{ferminum} we interpreted the $Z$ symmetry as imposing a
condition on the allowed fermion numbers of the states in the
theory. As the symmetry is broken in infinite volume, we conclude that
in infinite volume states with arbitrary fermion number may exist.
The screening state, being able to screen {\it any} charge, not only
multiples of the fundamental charge, is a non-local, polarized vacuum
like state, and as such it can a priori have any fermion number.


\section{Concluding Remarks}

In this paper we compute explicitly the effective action and the
correlators of the Polyakov loop operator in the one flavor Schwinger
model at finite temperature in order to investigate the phases of one
dimensional Q.E.D.  Our aim is to provide a convincing test of the
recent proposal \cite{gss} that the Polyakov loop operator is indeed
useful to distinguish between a confined and a deconfined phase of an
abelian model coupled with fermionic matter.

We present a form of the finite temperature effective action which
explicitly realizes the $Z$ symmetry.  We show that in one-dimensional
Q.E.D.  with massless fermions the $Z$ symmetry is not broken in
finite volume.  The $Z$-symmetry is broken --- due to strong infrared
effects --- only in the infinite volume limit where there is
off-diagonal long range order and the physical states may have
arbitrary fermion number.

In infinite volume, the thermal state of the Schwinger model can
screen an arbitrary external charge and, therefore, it is in the
deconfined phase.  Our explicit computation of the $Z$ symmetry
breaking in the Schwinger model is supported by two simple arguments
providing a sound physical intuition for the breaking of a discrete
symmetry in a one dimensional field theory. The massive Schwinger
model, on the other hand, is confining and the $Z$ symmetry is not
broken, at least when the temperature is much greater than the
electron mass and the confinement scale is set by the dimensional
electron charge.

\bigskip



\setlength{\baselineskip}{14pt}

\end{document}